\documentclass[%
 reprint,
 amsmath,amssymb,
 aps,
]{revtex4-2}
\usepackage{graphicx}
\usepackage{amsmath}
\usepackage{url}
\usepackage{float}
\usepackage[english]{babel}
\usepackage{chngcntr}
\usepackage{natbib}
\usepackage{textcase}
\usepackage{zref-xr}
\usepackage{hyperref}
\usepackage{tikz}
\usepackage{xcolor}
\usepackage{epstopdf}
\usepackage[percent]{overpic}
\usepackage{chngcntr}
\usepackage{mathtools}
\usepackage{rotating}
\usepackage{placeins}
\usepackage[absolute,overlay]{textpos}
\usepackage{array}

\begin{document}

\preprint{APS/123-QED}

\title{A comment on: ''Universal control of superexchange in linear triple quantum dots with an empty
mediator"}

\author{Marko J. Ran\v{c}i\'{c}$^1$}
 \email{marko.rancic@totalenergies.com}
\affiliation{TotalEnergies, 8, Boulevard Thomas Gobert, Nano-INNOV – Bât. 861,– 91120 Palaiseau – France }

\begin{abstract}
In a recent preprint \cite{chan2022universal}, G. X. Chan, P. Huang, and X. Wang claim that triple-quantum dot superexchange in a $(1,0,1)$ charge configuration exhibits a change of sign (going from positive to negative) as a function of middle dot detuning. Furthermore, their claim is that charge sweet-spots exist for specific values of the inter-dot detuning. Their analysis is based on the Hubbard model and something to what they refer to as the ''full Configuration-Interaction" method. All of this findings were already reported by M. J. Ran\v{c}i\'{c} and G. Burkard in Ref. \cite{PhysRevB.96.201304} (2017) based on the Hubbard model. No reference to this manuscript was made in Ref. \cite{chan2022universal}. I have asked the authors to urgently modify the pre-print and position their work with respect to the previously conducted study - which they rejected to do at the current moment, quoting that pp. ''it is not their style to modify preprints before they were accepted". This alongside with a very similar style of some figures lead me to the conclusion that they are deliberately misleading the scientific community and trying to adopt other peoples work and ideas as their own. 

\end{abstract}

\maketitle


\section{Introduction}
The triple quantum dot loaded with two electrons exhibits a plethora of rich physics. This system has been especially appealing for experimental studies in the area of semiconductor spin quantum information processing.
\section{M. J. Ran\v{c}i\'{c} and G. Burkard Phys. Rev. B 96, 201304 (2017)}
In Ref. \cite{PhysRevB.96.201304} starting from the Hubbard model without inter-dot Coulomb penalization

\begin{align}\label{eq:Ham}
 H=&\sum\limits_{i\sigma}(\varepsilon_i+E_{\rm z}^i{\bf \sigma})n_{i\sigma}+U\sum\limits_i n_{i\uparrow} n_{i\downarrow}+\sum\limits_{\langle ij \rangle}\sum \limits_{\sigma} t_{ij} c_{i\sigma}^{\dagger}c_{j\sigma},
\end{align} and applying the powerfull Schrieffer-Wolff transformation the following expression is derived for the superexchange interaction in the (1,0,1) charge configuration

\begin{equation}\label{eq:SupEx}
J_{\rm SE}=4 t^4 U\frac{U \left(12 \delta ^2+\epsilon ^2\right)-\delta  \left(8 \delta ^2+6 \epsilon ^2\right)}{\left(\epsilon ^2-4 \delta ^2\right)^2 \left(U-2 \delta \right) \left(U^2-\epsilon ^2\right)}.
\end{equation}
The expression in Eq. (\ref{eq:SupEx}) has contributions from six different super-exchange paths as seen in Tab. \ref{tab1}.

To complement on that, 4 charge noise sweet spots were derived - points in which the superexchange is simultaneously insensitive too changees in both the outer dot detuning $\epsilon$ and average-outer dot middle dot detuning $\delta$. 

Furthermore, analytical expressions for points in which superexchange is zero were derived

\begin{equation}\label{eq:delta0}
\delta_0=\frac{1}{2}\left(1+\frac{1-\epsilon^2}{q^{1/3}}+q^{1/3}\right);\,\epsilon_0=\pm\frac{2\sqrt{(3-2\delta)\delta^2}}{\sqrt{6\delta-1}},
\end{equation} where, $q=1-\epsilon^2+\sqrt{(\epsilon^2(\epsilon^2-1)^2)}$ all given in units of Coulomb repulsion U. this points are not simultaneously ''sweet spots" - superexchange changes signs in them.

\begin{center}
\begin{table*}[t]
\begin{tabular}{|c|c|c|c|}
\hline
 i & Superexchange path & Superexchange expression & Sign of $J_{\rm SE}^i$\\
 \hline

1 & ${(\uparrow,0,\downarrow)\xleftrightarrow[]{t_R}(\uparrow,\downarrow,0)\xleftrightarrow[]{t_L}(0,\uparrow\downarrow,0)\xleftrightarrow[]{t_L}(\downarrow,\uparrow,0)\xleftrightarrow[]{t_R}(\downarrow,0,\uparrow)}$ & $ t^4/\left[(U-2\delta)(\epsilon/2+\delta)^2\right]$  & $J^1_{\rm SE}>0$\\ 
2 & ${(\uparrow,0,\downarrow)\xleftrightarrow[]{t_L}(0,\uparrow,\downarrow)\xleftrightarrow[]{t_R}(0,\uparrow\downarrow,0)\xleftrightarrow[]{t_R}(0,\downarrow,\uparrow)\xleftrightarrow[]{t_L}(\downarrow,0,\uparrow)}$ & $ t^4/\left[(U-2\delta)(\epsilon/2-\delta)^2\right]$ & $J^2_{\rm SE}>0$\\ 
3 & ${(\uparrow,0,\downarrow)\xleftrightarrow[]{t_R}(\uparrow,\downarrow,0)\xleftrightarrow[]{t_L}(0,\uparrow\downarrow,0)\xleftrightarrow[]{t_L}(0,\uparrow,\downarrow)\xleftrightarrow[]{t_L}(\downarrow,0,\uparrow)}$ & $ -t^4/\left[(U-2\delta)(\epsilon/2-\delta)(\epsilon/2+\delta)\right]$ & $J^3_{\rm SE}<0$\\
4 & ${(\uparrow,0,\downarrow)\xleftrightarrow[]{t_L}(0,\uparrow,\downarrow)\xleftrightarrow[]{t_R}(0,\uparrow\downarrow,0)\xleftrightarrow[]{t_L}(\downarrow,\uparrow,0)\xleftrightarrow[]{t_R}(\downarrow,0,\uparrow)}$ & $ -t^4/\left[(U-2\delta)(\epsilon/2-\delta)(\epsilon/2+\delta)\right]$ & $J^4_{\rm SE}<0$\\ 
5 & ${(\uparrow,0,\downarrow)\xleftrightarrow[]{t_R}(\uparrow,\downarrow,0)\xleftrightarrow[]{t_L}(\uparrow\downarrow,0,0)\xleftrightarrow[]{t_L}(\downarrow,\uparrow,0)\xleftrightarrow[]{t_R}(\downarrow,0,\uparrow)}$ & $ t^4/\left[(U-\epsilon)(\epsilon/2+\delta)^2\right]$ & $J^5_{\rm SE}>0$\\ 
6 & ${(\uparrow,0,\downarrow)\xleftrightarrow[]{t_L}(0,\uparrow,\downarrow)\xleftrightarrow[]{t_R}(0,0,\uparrow\downarrow)\xleftrightarrow[]{t_R}(0,\downarrow,\uparrow)\xleftrightarrow[]{t_L}(\downarrow,0,\uparrow)}$ & $ t^4/\left[(U+\epsilon)(\epsilon/2-\delta)^2\right]$ & $J^6_{\rm SE}<0$ \\
\hline
\end{tabular}
\caption{Six possible superexchange paths involving spin-conserving tunneling with corresponding expressions $J_{\rm SE}=\sum_i J_{\rm SE}^i$. The parameters for which the sign of $J_{\rm SE}$ is valid are the Coulomb repulsion $U=1\text{ meV}$, the detuning between the outer dots $\epsilon=-1.34U$, the detuning between the middle dot and the average of the outer dots $-0.2U<\delta<0.3U$.}
\label{tab1}
\end{table*}
\end{center}

\section{G. Xu. Chan, P. Huang, and X. Wang arXiv:2203.15521 (2022)}

In much similarity to Ref. \cite{PhysRevB.96.201304} the authors base their study on the Hubbard model and the Schrieffer-Wolff transformation in the same (1,0,1) charge configuration. The sole difference is that they chose the inter-dot Coulomb penalization to be non-zero, leading to the fact that (1,1,0) and (0,1,1) charge states are raised higher in energy as compareed to (2,0,0), (0,2,0) and (0,0,2). This would correspondd to adding the following term in {Eq. (\ref{eq:Ham})} $\sum_{\langle ij \rangle}V n_in_j$. However, $U\gg V$ and neglecting $V$ in favor of $U$ simplifies calculations significantly.

\begin{figure*}[tbh!]
    \begin{minipage}{0.5\textwidth}
        \centering
        \includegraphics[scale=0.5]{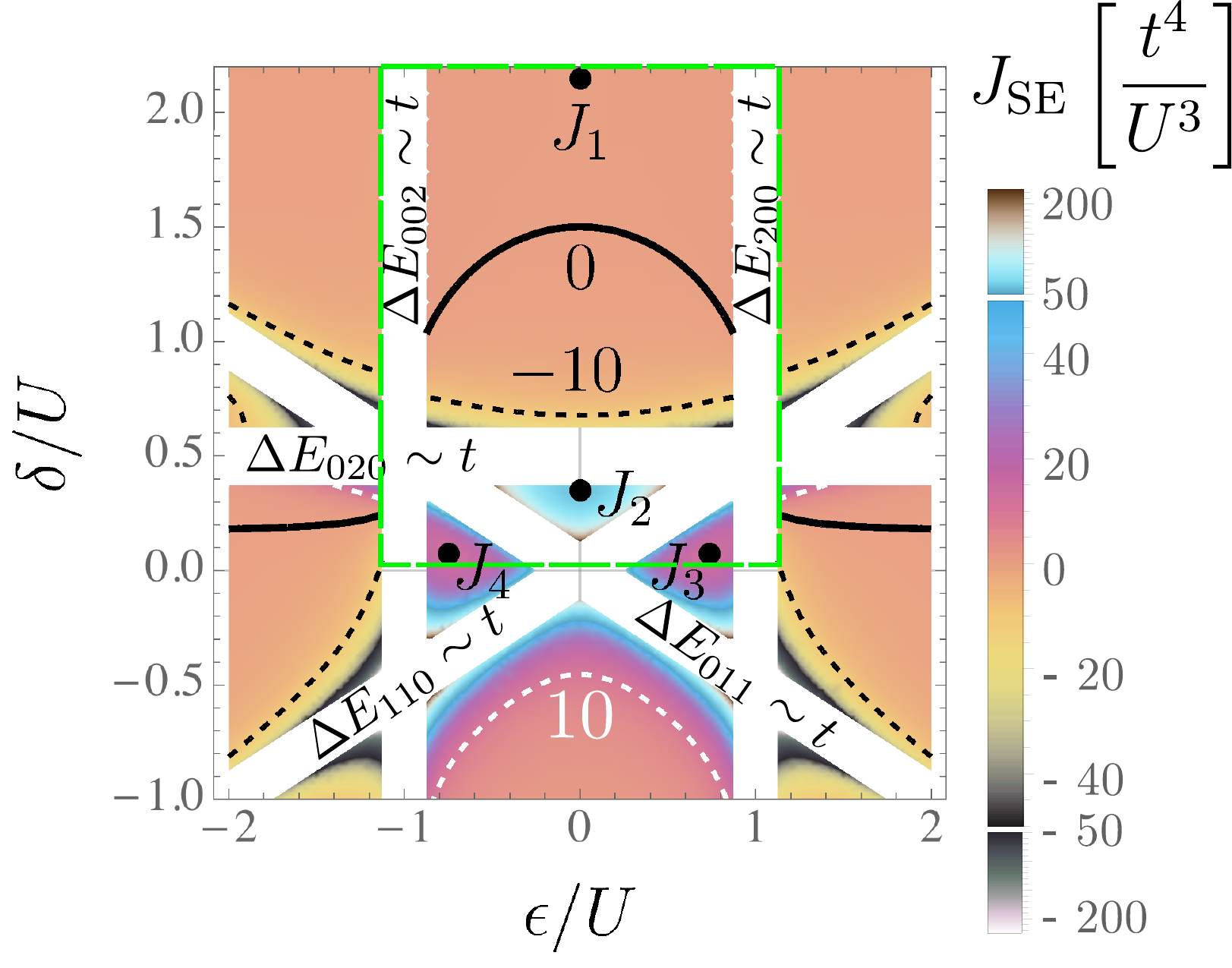}
        \put(-130,185){(a)}
        \put(75,185){(b)}
    \end{minipage}\begin{minipage}{0.5\textwidth}
        \centering
        \includegraphics[scale=0.5]{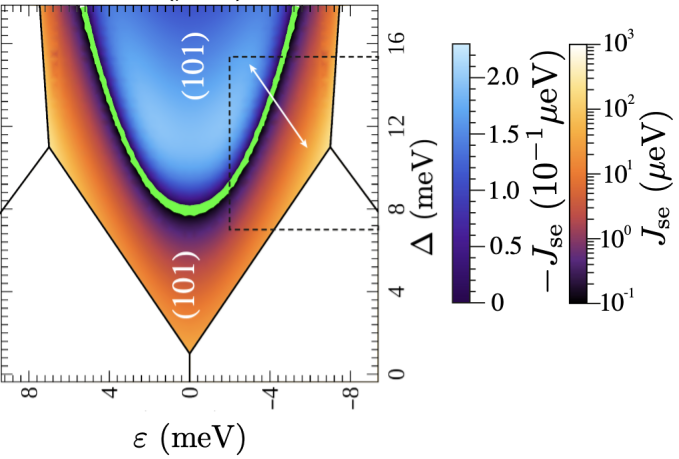}
    \end{minipage}
    \caption{Superexchange as a function of $\epsilon$ and $\delta$. (a) Results of Ref. \cite{PhysRevB.96.201304} and (b) results of \cite{chan2022universal}, figure rotated and legend rearranged so that the resemblance with subfigure (a) is obvious. Area to which authors  of Ref. \cite{chan2022universal} focus is marked with a green box in subfigure (a). \label{Fig:comp}}
\end{figure*}

The authors of Ref. \cite{chan2022universal} focus on a particular energy ordering in the part of the study (Section IIB) ${E_{101}<E_{110;011}<E_{200;002}}$, with all other states being treated as higher in energy. Such a process is described by $J_{\rm SE}^6$ in Tab. \ref{tab1}. By going through the calculations in Eqs. (3a-4c) I obtained a result $\sim J_{\rm SE}^6$, up to a numerical difference (factor of $\sqrt{2}$ the origin of which I cannot determine) and a factor of $2$ next to $\epsilon$ because different definitions were used.

In Fig. \ref{Fig:comp} a comparison betweeen Fig. 3 of Ref. \cite{PhysRevB.96.201304} (subfigure (a)) and Fig. 8 (f) of Ref. \cite{chan2022universal} is displayed (marked as subfigure (b)). The green area in subfigure (a) corresponds to that of subfigure (b) however is different with respect to subfigure (b) due to the presence of the (0,2,0) state. This is a consequence of the finite $V$ in the authors study which lowers the (0,2,0) state in energy relative to the (1,1,0) and (0,1,1). The concavity of $J_{\rm SE}=0$ line is also different an effect which could originate in finite and unrealistically large $V$ in authors study (however, given that I consider that their work is plagiarism the burden on understanding the differences of this two works lies on them).

The authors write in their abstract: ''We have found that, when the detunings at the left and right dots are leveled, the superexchange can exhibit a non-monotonic behavior which ranges from positive to
negative values as a function of the middle-dot detuning". While the conclusions of Ref. \cite{PhysRevB.96.201304} state: ''Furthermore, we have shown that the sign of the superexchange can be
changed by varying the detuning parameters". Same can bee seen by going along the line of $\epsilon=0$ in subfigure (a).

Finally, the authors claim to predict the existence of ''sweet spots" in Fig. 8 (d) and (h). Predictions of sweet spots has been the main goal of Ref. \cite{PhysRevB.96.201304}. They have been analytically found and are marked $J_1-J_4$ in Fig. \ref{Fig:comp}. 

Let me conclude that in my opinion some scientific value might exist in Ref. \cite{chan2022universal} in the full Configuration-Interaction numerical calculations of superexchange which the authors claim to have done. However, no level of detail is provided about details of such calculations.

To sum up, Ref. \cite{chan2022universal} copies all findings of Ref. \cite{PhysRevB.96.201304} without referencing to it. The authors have rejected to immediately modify their pre-print and position their work with respect to a previous study which has been accepted in a prestigeous peer-reviewed journal in 2017. In my opinion some scientific value might exist in the full Configuration-Interaction calculations the authors claim to have done but no detail about them is provided. Such work would clearly have be differentiated from Ref. \cite{PhysRevB.102.035427} if it was to have any value what-so-ever.

\bibliography{bibli}

\end{document}